 \documentclass[preprint2,longabstract]{aastex}

\usepackage[varg]{txfonts}
\usepackage{graphicx}
\usepackage{amssymb}
\usepackage{lscape}
\usepackage{multirow}
\usepackage{color}
\usepackage{longtable}
\bibpunct{(}{)}{;}{a}{}{,}

\def\logg {\log g}

\def\ds {$\delta$ Sct}
\def\dss {$\delta$ Sct stars}

\def\corot {{CoRoT}}

\def\kepler {{\it{Kepler}}}

\def\logg {$\log g$}

\def\vsini {$\mathrm{v}\cdot\!\sin\!~i$}
\def\kms {$\mathrm{km}~\mathrm{s}^{-1}$}
\def\teff {$T_{\mathrm{eff}}$}

\def\muhz {$\mu\mbox{Hz}$}

\def\cd {$\mbox{d}^{-1}$}

\def\Dnu {$\Delta\nu$}

\def\rhom {$\bar\rho$}

\def\ele{$\ell$}

\def\Ok{$\Omega_{\mathrm{K}}$}
\def\stara {KIC\,3858884}
\def\starb {KIC\,4544587}
\def\starc {KIC\,10661783}
\def\stard {HD\,172189}
\def\stare {CID\,100866999}
\def\starf {CID\,105906206}
\def\starg {HD\,159561}

\def\gh {Garc\'{\i}a Hern\'andez}
\slugcomment{{\sc Accepted to ApJL:} September 3, 2015}

\shorttitle{Observational \Dnu-\rhom\ relation for \dss}
\shortauthors{Garc\'{\i}a Hern\'andez et al.}

\begin{document}


\title{Observational \Dnu-\rhom\ relation for \dss\ using eclipsing binaries and space photometry}
%\title{Frequency patterns as the key to unlock the secrets of pulsations in \dss}


\author{A. \gh}
\affil{Instituto de Astrof\'{\i}sica e Ci\^{e}ncias do Espaço, Universidade do Porto, CAUP, Rua das Estrelas, PT4150-762 Porto, Portugal}
\email{agh@astro.up.pt}

\author{S. Mart\'{\i}n-Ruiz}
\affil{Instituto de Astrof\'{\i}sica de Andaluc\'{\i}a (CSIC), Glorieta de la Astronom\'{\i}a S/N, 18008, Granada, Spain}

\author{M\'ario J. P. F. G. Monteiro}
\affil{Instituto de Astrof\'{\i}sica e Ci\^{e}ncias do Espaço, Universidade do Porto, CAUP, Rua das Estrelas, PT4150-762 Porto, Portugal}
\affil{Departamento de F\'{\i}sica e Astronomia, Faculdade de Ci\^{e}ncias da Universidade do Porto, Rua do Campo Alegre, 4169-007 Porto, Portugal}

\author{J. C. Su\'arez}
\affil{Universidad de Granada. Dept. F\'{\i}sica Te\'orica y de Cosmos. Fuentenueva Campus. 18007. Granada, Spain}
\affil{Instituto de Astrof\'{\i}sica de Andaluc\'{\i}a (CSIC), Glorieta de la Astronom\'{\i}a S/N, 18008, Granada, Spain}

\author{D. R. Reese}
\affil{School of Physics and Astronomy, University of Birmingham, Edgbaston, Birmingham, B15 2TT, UK}
\affil{Stellar Astrophysics Centre (SAC), Department of Physics and Astronomy, Aarhus University, Ny Munkegade 120, DK-8000 Aarhus C, Denmark}

\author{J. Pascual-Granado}
\affil{Instituto de Astrof\'{\i}sica de Andaluc\'{\i}a (CSIC), Glorieta de la Astronom\'{\i}a S/N, 18008, Granada, Spain}

\and

\author{R. Garrido}
\affil{Instituto de Astrof\'{\i}sica de Andaluc\'{\i}a (CSIC), Glorieta de la Astronom\'{\i}a S/N, 18008, Granada, Spain}


%\altaffiltext{1}{Visiting Astronomer, Cerro Tololo Inter-American Observatory.
%CTIO is operated by AURA, Inc.\ under contract to the National Science
%\altaffiltext{2}{Society of Fellows, Harvard University.}
%\altaffiltext{3}{present address: Center for Astrophysics,
%    60 Garden Street, Cambridge, MA 02138}
%\altaffiltext{4}{Visiting Programmer, Space Telescope Science Institute}
%\altaffiltext{5}{Patron, Alonso's Bar and Grill}


\begin{abstract}
Delta Scuti (\ds) stars are intermediate-mass pulsators, whose intrinsic oscillations have been studied for decades. However, modelling their pulsations remains a real theoretical challenge, thereby even hampering the precise determination of global stellar parameters. In this work, we used space photometry observations of eclipsing binaries with a \ds\ component to obtain reliable physical parameters and oscillation frequencies. Using that information, we derived an observational scaling relation between the stellar mean density and a frequency pattern in the oscillation spectrum. This pattern is analogous to the solar-like large separation but in the low order regime. We also show that this relation is independent of the rotation rate. These findings open the possibility of accurately characterizing this type of pulsator and validate the frequency pattern as a new observable for \dss.
\end{abstract}





\section{Introduction}

\dss\ are intermediate-mass, main-sequence stars with particularly interesting internal physical properties. After white dwarfs, \dss\ are the second most numerous pulsators in the Galaxy \citep{Breger1979}. Although considered as potential ``standard candles", their complex oscillation spectra are dominated by non-radial modes, which hamper the identification of modes, and hence the period-luminosity relation.\par

Classical theories on stellar interiors and pulsations are not of much help in our understanding of the pulsation spectra of these stars. \dss\ are A-F type stars, commonly known to be fast rotators \citep{Royer2007}. The centrifugal force deforms the stellar structure to such a point that 1D asteroseismic modeling fails to correctly describe their oscillations.\par

Nonetheless, various efforts have been made to overcome these issues. Periodic patterns within the frequency sets of \dss\ have been systematically searched for from the ground with partial success \citep{Handler1997, Breger1999}. In the era of high-precision photometry from space missions like \corot\ \citep{corot} and \kepler\ \citep{kepler}, such patterns have been detected more successfully and characterized \citep[hereafter GH09, GH13a and GH13b, respectively]{GH09, GH13a, GH13b}. However, the connection between this periodicity and stellar properties has never been established with observations of \ds\ frequencies.\par

This is due to the fact that \dss\ pulsate around the fundamental radial mode (i.e., low order regime), where no obvious spacings are expected. The opposite case is represented by solar-like pulsators, whose pressure modes ($p$ modes) fall in the so-called asymptotic regime, where $n\gg\ell$. A pattern is clearly visible there, forming the large frequency separation (\Dnu), an almost constant spacing between modes of consecutive radial orders ($n$) of the same spherical degree (\ele). A scaling relation can also be obtained, relating \Dnu\ with the mean density of the star (\rhom) in a simple way \citep{Tassoul1980}. Although $p$ modes are expected to scale, in general, with the mean density of the star, other effects, such as rotation, will change the value and number of pulsation frequencies. Therefore, a direct dependency of the frequencies on the mean density requires validation, when using low order modes of \dss.\par

From a theoretical point of view, various works have also pointed out that this pattern could indeed be linked to the mean density in a simple way. For instance, a recent theoretical work using the most advanced 2D pulsating models suggested that a large separation within the $p$ modes of rapid rotators could be related in a simple way to the mean density of the star \citep{Reese2008}. Moreover, that relation is the same at any rotation rate. A very recently published work by \citet{Ouazzani2015} used realistic distorted pulsating models to build Echelle diagrams. They showed that a large separation structure formed by the so-called {\emph{island modes}} \citep{Lignieres2006} appeared in the rapid rotation case. Moreover, similarly to the solar-like relation, \citet{Lignieres2008, Lignieres2009} showed that this value is proportional to the inverse of the sound travel time along the underlying ray path. Other studies, using a grid of non-rotating models, found an equivalent scaling relation between \Dnu\ of \dss, i.e., in the low order regime, and \rhom\ of the corresponding model \citep{Suarez2014}.\par

In this work, we provide an empirical counterpart to the previous results, by studying a sample of eclipsing binary stars for which patterns have been detected. Reliable stellar densities obtained from binarity are compared with those predicted by patterns. Additionally, we needed these systems to be observed with high-precision photometry, so we only selected systems observed by space missions such as MOST \citep{most}, \corot\ \citep{corot} and \kepler\ \citep{kepler}. In what follows, we confirm all the theoretical results and compare our findings with the solar-like scaling relation.\par

\section{The sample}

The observational validation of the \Dnu-\rhom\ relation for \dss\ needs to be independent of stellar interior or pulsation modelling. To achieve such an objective, we searched for and selected eclipsing binary systems with a \ds\ component. Eclipsing binaries allow us to determine the mass and the radius of both components by means of geometrical considerations, Kepler's Third Law, and stellar atmosphere models. These significantly reduce the effects of rotation on inferred properties. Using the mass and the radius of eclipsing binaries, we were able to obtain a model-independent determination (in terms of stellar interior and evolution theories) of the mean density of the pulsating component.\par

The second variable needed to derive a large separation-mean density relation is the large separation contained within the frequency spectrum of the \ds\ component. We followed the method described in GH09 and GH13a to derive this value. As in the case analyzed by these authors, space observations were needed to ensure a statistically significant number of frequencies. This does not necessarily imply a huge number, but one that is sufficient to enable us to find a quasiperiodic behaviour. Consequently, the selected sample is composed of systems observed with the MOST, \corot\ and \kepler\ satellites.\par

We found a total of 7 systems fulfilling the above conditions: 6 eclipsing binaries and a binary for which the main \ds\ component was resolved using optical interferometry. The relevant information used in this work, mainly the mass, the radius, the large separation, and the corresponding references are summarized in Table~\ref{tab:sample}.\par

\section{Frequencies selection}
\label{Sec:Frequencies}

The effectiveness of the Fourier transform method at detecting the pattern in the frequency sets depends on the reliability of the frequencies used. We need to avoid frequencies that do not correspond to true intrinsic pulsation modes of the star, such as harmonics or combinations. We also need to restrict our analysis to the $p$ mode region. To that end, we cleaned the frequency sets before carrying out our analysis.\par

All of the stars in the sample showed two well separated domains in their frequency sets. The set at lower frequencies was identified as the $g$ mode region, whereas the other set as the corresponding $p$ mode range. We only used frequencies lying within the $p$ mode region in our search for periodicities.\par

Only \starg\ ($\alpha$~Oph) did not show a clear separation between $p$ and $g$ modes \citep{Monnier2010}. Instead, the authors identified a region in the low order regime, up to 193~\muhz\ (12 \cd), with frequencies clustered around certain values. They interpreted the separation between consecutive clusters as the rotational splitting and the modes in this regime as a set of $g$ modes. \citeauthor{Monnier2010} also estimated the fundamental radial mode as $\sim116$~\muhz\ (10~\cd). We thus avoided frequencies below this limit and those marked as probable gravity modes in our analysis. This enhanced the peak of the periodic pattern in the Fourier analysis (see Sec.~\ref{Sec:FT}). In any case, the frequency pattern is also detectable, although with some difficulty, when the complete set is used.\par

There was a large variety in the number of detected frequencies for each pulsators. The stars \stara\ \citep{Maceroni2014}, \stare\ \citep{Chapellier2013} and \starf\ \citep[where CID represents the \corot\ ID number]{daSilva2014} showed hundred of frequencies. But the most populated spectrum was shown by \starg\ \citep{Monnier2010}, with almost two thousand frequencies.\par

Nonetheless, after subtracting harmonics and combinations, as well as removing gravity modes, the number of independent modes decreased drastically. Two stars, \stara\ and \starf, remained with more than two hundred frequencies. We note that, even in those cases, the frequency pattern appeared clearly when using the subsets of the 30 and 60 highest-amplitude peaks for the Fourier analysis (see GH09 and GH13a).\par

For this reason, we only used the 50 highest amplitude modes in the case of \stard\ (see Appendix). All these frequencies, except $\mathrm{f_{25}}=3.91$~\muhz\ (0.3378~\cd), are above 160~\muhz\ ($\sim$14~\cd). Thus, we considered them as pressure modes.\par

\section{Obtaining the frequency pattern}
\label{Sec:FT}

Obtaining the large separation in the low order regime is a complex part of the analysis. We do not expect to see a large number of modes well aligned in the frequency spectrum (GH09), as in the case of solar-like stars. We performed the analysis described in GH09 and GH13a for the detection of the periodicity. In addition, we built Echelle diagrams which allowed us to confirm the presence of a pattern, measure \Dnu\ with high accuracy ($\pm1$~\muhz\ = 0.0864~\cd) and disentangle different possibilities in the most complicated cases (GH13a).\par

It is expected that in certain cases periodicities appear as a submultiple of the large separation (see GH09 and GH13a). The cases where half of \Dnu\ is found, i.e., for \stard\ and \starg, were expected from theory \citep{Pasek2012}. The moderate rotator \starb\ shows $\Delta\nu/4$ as the most prominent peak in the Fourier transform analysis. Interestingly, this peak corresponds to twice the rotational splitting. Estimates of the rotational splittings were obtained using the projected equatorial rotation velocity. We interpreted the radius derived from the binary analysis of the system as the equatorial radius, and the inclination angle of the system as the inclination of the rotational axis (see Sec.~\ref{Sec:Rotation}). \par

Although \starf\ is a {\it slow} \ds\ rotator, its oscillation spectrum presented more difficulties in the analysis than the other stars in the sample. The reason for that could be the evolved state of this star, as inferred from its fundamental parameters. An evolved star will present avoided crossings, blurring the large separation pattern. For this reason, it has a 2\,\muhz\ (= 0.1728~\cd) uncertainty in the value of \Dnu\ listed in Table~\ref{tab:sample}.\par

\section{Calculating the mean density}

Since we are dealing with objects affected by the centrifugal force, we tried to avoid the spherical approximation in all of our calculations. That is why we used the Roche model approximation \citep[e.g.][]{Maeder2009} to compute the mean density of each star, assuming that the measured radii correspond to the equatorial ones. The exception is \starg, for which we have the equatorial and polar radii.\par

Ideally, one would want to calculate the mean densities of realistic stellar models. However, for the sake of simplicity, it is useful to consider the mean density of a Roche model. Such an approximation was validated through comparisons with more realistic Self-Consistent Field (SCF) models \citep{MacGregor2007}. Hence, they provide a reliable estimate of the geometric properties of $\delta$ Scuti stars, as long as their rotation rate is close to uniform.\par

The uncertainties in the mass and radius allowed us to calculate the errors in the mean densities. A standard error propagation method was used to compute them. We also assumed that the uncertainties in the equatorial and polar radii were the same.\par

\section{The \Dnu-\rhom\ relation}

\begin{figure}
\epsscale{.95}
\plotone{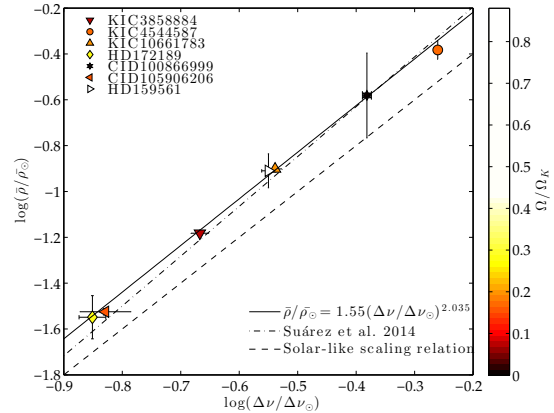}
\caption{Large separation-mean density relation obtained for the 7 binary systems of our sample. A linear fit to the points is also depicted, as well as the solar-like scaling relation \citep{Tassoul1980} and the theoretical scaling relation for non-rotating models of \dss\ \citep{Suarez2014}. Symbols are plotted with a gradient colour scale to account for the different rotation rates.\label{fig:Dnu_rho}}
\end{figure}

In Fig.~\ref{fig:Dnu_rho}, we plotted all the information extracted for the sample, namely, \Dnu, \rhom\ and $\Omega$, in a $\log\Delta\nu$ versus $\log\bar\rho$ diagram and indicated the rotation rate via the colours of the symbols. The rotation rates were computed using the information from binary analysis (see Sec.~\ref{Sec:Rotation}). These symbols follow a clear linear trend. A weighted linear fit, taking uncertainties as weights, gave the following expression:\par

\begin{equation}
\bar\rho/\bar\rho_{\odot}=1.55^{+1.07}_{-0.68}(\Delta\nu/\Delta\nu_{\odot})^{2.035\pm0.095}.
\end{equation}

A normalization by the solar values \citep[with $\Delta\nu_{\odot}$ = 134.8 \muhz,][]{Kjeldsen2008} is useful because it allows us to compare this trend with the scaling relations from \citet{Suarez2014}, and \citet{Tassoul1980} for solar-like pulsators, as depicted in Fig. 1.The uncertainties correspond to the 95\%\ confidence bounds on the coefficients of the linear fit.\par

To within the uncertainties, the slope of the linear trend fits the results found with non-rotating models, but this trend is offset by a multiplicative constant in comparison with the solar-like scaling relation. This was expected since the regime of \ds\ pulsations is different from the latter, but the agreement with the non-rotating case is the sign of the invariance of the relation with rotation as predicted by \citet{Reese2008}.\par

Thus Fig.~\ref{fig:Dnu_rho} confirms several hypotheses. The main one is that the large frequency separation calculated in the low order regime (relevant to \dss) is still related in a simple way to the mean density of the star, thereby enabling us to determine this quantity without additional information. The second consequence is that this relation does not depend on the rotation rate as clearly shown in the plot and discussed in detail in the following section.\par

\section{The independence of the \Dnu-\rhom\ relation respect to rotation}
\label{Sec:Rotation}

The binarity allowed us to estimate the rotation rate of the star. We assumed the rotation axis to be perpendicular to the orbital plane of the binary system, which is adequate in most cases. Using the observed values for the radius and the projected equatorial rotation velocity, \vsini, with $i$ being the inclination of the rotation axis with respect to the observer, we computed the rotation rate, $\Omega$. We expressed the results in units of \Ok, \Ok\ being the Keplerian break-up rotation rate, which is approximately the maximum velocity for which the star does not lose mass due to the centrifugal acceleration. We did not include uncertainties in the $\Omega$ values, since they are only estimates and do not alter the conclusions.\par

Computing the rotation rate allowed us to study the impact of rotation on the \Dnu-\rhom\ relation. This is an important point, since A-type stars are usually fast rotators, as stated earlier. Indeed, our sample covers a wide range from 0.0754 to 0.88\Ok, this upper limit being reached by \starg\ \citep{Monnier2010}. An invariance of the trend with respect to rotation rate will mean that \Dnu\ is a really useful new observable for \dss.\par

\begin{figure}
\epsscale{.95}
\plotone{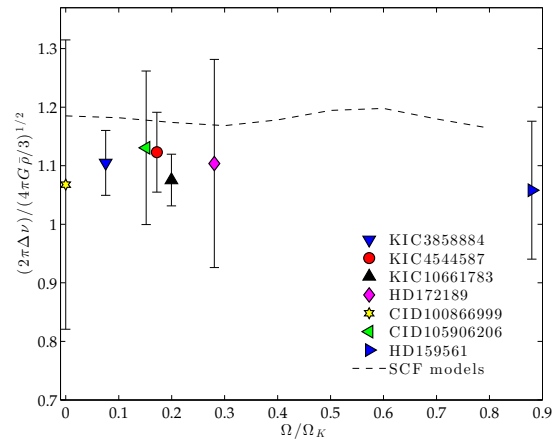}
\caption{Large frequency separation as a function of the rotation rate. \Dnu\ has been scaled by the square root of the mean density. The relation found using $2~\mathrm{M}_\odot$ SCF models is also plotted as a reference. \label{fig:Dnu_Ohm}}
\end{figure}

This invariance was already theoretically predicted by \citet{Reese2008}. For a series of 2D rotating polytropic models with $\Omega$/\Ok\ = [0, 0.6], they showed that \Dnu\ is related to the mean density and that the ratio \Dnu/\rhom\ is almost constant at any rotation. We wanted to compare our findings with those theoretical predictions. To that end, we have reproduced Fig.~2 of \citet{Reese2008} but using a new set of $2~\mathrm{M}_\odot$ SCF models \citep{Reese2009} instead of polytropic ones, and adding the results of the analysis presented here. The final plot can be seen in Fig.~\ref{fig:Dnu_Ohm}.\par

The SCF models were computed up to 0.8~\Ok, so they did not account for the high rotation rate of \starg. In any case, it is clear that an almost constant trend is preserved. In addition, one may notice that our sample is grouped around a value of 1.1 in the y-axis, instead of 1.2 found in the modelling. One possible explanation is that the set of modes used in the theoretical calculations have higher radial orders than in the stars. It is extremely time consuming to look for low order $p$ modes in the theoretical calculations as they tend to get lost in a sea of g-modes.\par

Nonetheless, the important conclusion of the conservation of the \Dnu-\rhom\ relation with rotation does not change. This implies that non-rotating models might be useful to estimate not only the mean density of the star but, in some cases, other characteristics, such as the evolutionary stage. Simple comparisons with non-rotating stellar evolution models, taking into account the usual observables, effective temperature, surface gravity and metallicity (\teff, \logg, [Fe/H]), as well as \Dnu\ from this work, provide such information \citep{Hareter2014}. This is not a measure of the age but an indicator of the stage within the evolutionary sequence of the star. For instance, the low \Dnu\ value of \starf\ and \stard\ could be the indicative of an evolved state, as it is also suggested by their fundamental parameters.\par

\section{Conclusions}

In this work, we present a scaling relation between a frequency pattern in the pulsation spectra of \dss\ (equivalent to the large frequency separation in solar-like stars) and their mean density. We used eclipsing binaries in which one of the components is a \ds\ pulsator to precisely determine the mass and the radius of the stars in the system, and satellite observations to derive the frequency pattern in the pulsating component. Our result is formally the same as the theoretical relation found by \cite{Suarez2014} using non-rotating models. Indeed, \citeauthor{Suarez2014}'s relation lies within our 1-$\sigma$ error bars.\par

Our scaling relation is analogous to that found in solar-like pulsators apart from a multiplicative offset. This difference could be due to the different pulsating regimes in which the periodicity is identified. For the large separation it is the asymptotic regime, whereas for \dss\ it is the low radial order region, closer to the fundamental radial mode.\par

From the mean density it is possible to identify the fundamental radial mode of the star. This result, together with the Echelle diagram and comparisons with realistic 2D rotating models may allow us to identify pulsation modes. Such an approach represents a first step towards a reliable mode identification method for \dss.\par

The second prominent result we obtained is that the relation does not depend on the rotation rate of the star. Previous theoretical works predicted the approximate invariability of the relation with rotation \citep{Reese2008}. The observational \Dnu-\rhom\ relation validates these results as well as those based on more realistic 2D models.\par

Such invariance implies that the mean density could be accurately determined for any \ds\ star, independently of its rotation rate (at least, up to $\Omega$ = 0.88~\Ok, the highest value in the present sample). Additional information might then be derived from this parameter. If a well-determined luminosity exists, a mass-luminosity relation \citep{Ibanoglu2006} would give the mass of the star. Once the mass is found, the radius of the star might be estimated. Using the derived luminosities of the stars from our sample, we tested the \citeauthor{Ibanoglu2006}'s relation. For the main-sequence stars, the agreement on the mass was found to be within 10\%. The Gaia mission \citep{deBruijne2012} will also play an important role in obtaining better mass estimates.\par

More observations are required to increase the size of the sample. Future space missions \citep[such as PLATO 2.0,][]{Rauer2014} and ground-based asteroseismic observations will help us to refine the present \Dnu-\rhom\ relation and thus provide mean densities with better accuracy. Also, a larger sample will allow us to understand better how rotation is related to other periodicities found in the oscillation spectrum.\par

\acknowledgments

AGH acknowledges support from Funda\c{c}\~ao para a Ci\^{e}ncia e a Tecnologia (FCT, Portugal) through the fellowship SFRH/BPD/80619/2011. AGH and JCS acknowledges support from the EC Project SPACEINN (FP7-SPACE-2012-312844). JCS also acknowledges funding support from the Spanish ``Ministerio de Econom{\'{\i}}a y Competitividad'' under ``Ram\'on y Cajal'' subprogram. DRR is currently funded by the European Community's Seventh Framework Programme (FP7/2007-2013) under grant agreement no.~312844 (SPACEINN), which is gratefully acknowledged. This research made use of the SIMBAD database and the VizieR catalogue access tool operated at CDS, Strasbourg, France, and the SAO/NASA Astrophysics Data System.

%We are grateful to V. Barger, T. Han, and R. J. N. Phillips for
%doing the math in section~\ref{bozomath}.
%More information on the AASTeX macros package is available \\ at
%\url{http://www.aas.org/publications/aastex}.
%For technical support, please write to
%\email{aastex-help@aas.org}.

\appendix

\section{Pulsation frequencies of \stard}
\label{Sec:Appendix}

The system \stard\ (CID~8170) was observed by the \corot\ satellite during the Second Long Run (LRc02). It was discovered by \citet{Martin2003} and was studied as a detached binary system with a \ds-type pulsating component by \citet{MartinRuiz2005}.\par

We used a new gap-filling method that is based on an autoregressive moving average interpolation \citep{PascualGranado2015} to remove the spectral window introduced by wrong data due to the South Atlantic Anomaly. This method recovers the signal as much as possible, allowing to improve the frequency determination. In addition, the curve was corrected for a decreasing trend.\par

The light curve has been analyzed using the SigSpec code \citep{Reegen2007}. We also investigated harmonics and combinations between the main peaks, as well as those related to the satellite orbital frequency $\mathrm{f_{s}}=161.713$~\muhz\ (13.972~\cd).\par

The frequencies used in this study are listed in Table~\ref{tab:freqs} together with the most relevant parameters. Errors for the frequencies, amplitudes and phases are estimated using the approximation given in \citet{Montgomery1999}.\par




%-table-%
\begin{landscape}
\begin{table}[p] % Special place 'p' does not work
 \caption{Characteristics of the systems taken from the literature (see footnote for references). The information corresponds to the pulsating component (which is not necessarily the primary).}
 \begin{footnotesize}
 \begin{center}
 \begin{tabular}{ccccccccc}
  \hline
  \hline
  System & \Dnu\ (\muhz) & M (M$_\odot$) & R (R$_\odot$) & \rhom\ $(\rho_{\odot})$ & \vsini\ (\kms) & i ($\circ$) & $\Omega/\Omega_{\mathrm{K}}$ \\
  \hline
  \stara$^1$ & $29\pm1$ & $1.86\pm0.04$ & $3.05\pm0.01$ & $0.0657\pm0.0021$ & $25.7\pm1.5$ & $88.176\pm0.002$ & 0.0754  \\
  \starb$^2$ & $74\pm1$ & $1.61\pm0.06$ & $1.58\pm0.03$ & $0.414\pm0.039$ & $75.8\pm15$ & $87.9\pm3$ & 0.172  \\
  \starc$^3$ & $39\pm1$ & $2.100\pm0.028$ & $2.575\pm0.015$ & $0.1255\pm0.0039$ & $78\pm3$ & $82.39\pm0.23$ & 0.200 \\
  \stard$^4$ & $19\pm1$ & $1.78\pm0.24$ & $4.03\pm0.11$ & $0.0283\pm0.0061$ & $78\pm3$ & $73.2\pm0.6$ & 0.281 \\
  \stare9$^5$ & $56\pm1$ & $1.8\pm0.2$ & $1.9\pm0.2$ & $0.262\pm0.112$ & -- & $80\pm2$ & -- \\
  \starf$^6$ & $20\pm2$ & $2.25\pm0.04$ & $4.24\pm0.02$ & $0.02986\pm0.00095$ & $47.8\pm0.5$ & $81.42\pm0.13$ & 0.152  \\
  \multirow{2}{*}{\starg$^7$} & \multirow{2}{*}{$38\pm1$} & \multirow{2}{*}{$2.40^{+0.23}_{-0.37}$} & $\mathrm{R}_\mathrm{eq}=2.858\pm0.015$ & \multirow{2}{*}{$0.123\pm0.021$} & \multirow{2}{*}{$239\pm12$} & \multirow{2}{*}{$87.5\pm0.6$} & \multirow{2}{*}{0.88}  \\
   &  &  & $\mathrm{R}_\mathrm{pol}=2.388\pm0.013$ &  &  &  \\
  \hline
  \hline
 \end{tabular}
 \label{tab:sample}
 \end{center}
 {$^1$\citet{Maceroni2014}; $^2$\citet{Hambleton2013}; $^3$\citet{Lehmann2013}; $^4$\citet{Creevey2009}; $^5$\citet{Chapellier2013}; $^6$\citet{daSilva2014}; $^7$\citet{Monnier2010, Hinkley2011}.}
 \end{footnotesize}
\end{table}
\end{landscape}
%-table-%

\begin{deluxetable}{cccccccc}

\tabletypesize{\scriptsize}
%\tabletypesize{\small}

\tablecaption{\small{The frequencies of the fifty highest amplitude modes found in the light curve of \stard.} \label{tab:freqs}}
\tablewidth{0pt}

\tablehead{%
 \colhead{N$^{\circ}$} &  \colhead{Frequency (\cd)} & \colhead{Frequency (\muhz)} & \colhead{Amplitude (ppm)} & \colhead{Phase (rad)} &  \colhead{sig} &  \colhead{S/N} & \colhead{rms}
}
 
\startdata
  1 & 19.5831716(39)      &  226.657079(45) & 722.6 (8) & 1.08676  (17) & 14367.0 &  305.1 & 1273.473 \\
  2  & 17.3221220(65)    &  200.487523(75) & 434.6 (8) & 4.65814  (28) & 11291.3 &  174.2 & 1001.383  \\
  3  & 17.9439181(67)     & 207.684237(78) & 418.3 (8) & 1.98871  (29) & 9342.6 &  161.5 &  934.435  \\
  4  & 17.8484710(69)     & 206.579525(80) & 409.4 (8) & 3.44651  (30) & 10557.7 &  159.0 &  831.672    \\
  5  & 19.4787320(72)     & 225.448287(83) & 392.7 (8) & 1.02567  (31) & 6436.4 &  164.1 &  746.122    \\
  6  & 18.0235161(91)     & 208.605510(105) & 309.0 (8) & 4.01698  (39) & 6907.3 &  118.9 &  718.091  \\
  7  & 17.8987022(138)    & 207.160905(160) & 203.7 (8) & 2.30491  (59) & 3973.1 &   79.0 &  669.483  \\
  8  & 17.0218038(138)    & 197.011618(160) & 203.5 (8) & 2.07516  (59) & 4155.8 &   80.7 &  654.006  \\
  9  & 18.8269103(138)    & 217.904054(160) & 203.6 (8) & 2.07137  (59) & 4286.3 &   81.7 &  638.259  \\
  10 & 18.8817741(172)   & 218.539052(199) & 163.7 (8) & 1.39528  (74) & 3329.0 &   65.9 &  608.236  \\
  11 & 18.6455834(161)   & 215.805363(186) &  174.2 (8) & 0.58922  (69) & 3235.2 &   68.7 &  596.494 \\
  12 & 15.3526920(524)   & 177.693194(607) &  53.6 (8) & 6.10732 (226) & 1837.3 &   21.7 &  557.451 \\
  13 & 15.5902540(242)   & 180.442755(280) & 116.1 (8) & 1.48105 (104) & 1826.1 &   46.7 &  551.558\\
  14 & 18.6002092(241)   & 215.280199(279) & 116.7 (8) & 0.90201 (104) & 1842.2 &   45.9 &  539.979\\
  15 & 18.4248333(267)   & 213.250385(309) & 105.4 (8) & 2.65838 (115) & 1767.4 &   41.4 &  534.238\\
  16 & 18.1590787(279)   & 210.174522(323) & 100.6 (8) & 3.92877 (120) & 1618.4 &   39.1 &  523.636\\
  17 & 17.8212096(329)   & 206.264000(381)  & 85.3 (8) & 0.14432 (142) & 1250.1 &   33.4 &  509.685 \\
  18 & 15.4156486(183)   & 178.421859(212)  & 153.3 (8) & 5.50644 (079) & 1146.3 &   61.7 &  499.145\\
  19 & 17.2956279(376)   & 200.180878(435) &  74.6 (8) & 0.87827 (162) &  988.5 &   29.9 &  487.064 \\
  20 & 19.5284410(403)   & 226.023623(466)  & 69.6 (8) & 1.08476 (174) &  943.6 &   29.3 &  481.494 \\
  21 & 24.4302374(411)   & 282.757377(476) &  68.4 (8) & 3.87273 (177) &  844.0 &   31.9 &  476.268 \\
  22 & 20.2295095(423)   & 234.137841(490) &  66.5 (8) & 2.81981 (182) &  839.1 &   28.5 &  473.961 \\
  23 & 17.0670180(452)   & 197.534931(523) &  62.2 (8) & 4.95846 (194) &  720.5 &   24.6 &  469.518 \\
  24 & 18.5500094(473)   & 214.699183(548) &  59.3 (8) & 2.24606 (204) &  721.9 &   23.3 &  467.571 \\
  25 & 0.3377559(534)     &  3.909212(618) &  52.6 (8) & 2.22065 (230) &  657.5 &   16.9 &  463.801 \\
  26 & 16.2914539(527)   & 188.558494(610) &  53.3 (8) & 0.58305 (227) &  607.3 &   21.2 &  458.742 \\
  27 & 17.7347792(514)   & 205.263648(595)  & 54.7 (8) & 3.85435 (221) &  525.5 &   21.5 &  455.685 \\
  28 & 20.5208666(612)   & 237.510030(708)  & 45.9 (8) & 4.20510 (263) &  494.0 &   19.9 &  450.248 \\
  29 & 22.9908392(640)   & 266.097676(741)  & 43.9 (8) & 0.21879 (275) &  428.4 &   20.0 &  442.027 \\
  30 & 15.1298017(724)   & 175.113446(838)  & 38.8 (8) & 1.73935 (312) &  361.2 &   15.9 &  436.897 \\
  31 & 14.8250206(756)   & 171.585887(875)  & 37.2 (8) & 5.74037 (325) &  358.7 &   15.3 &  435.991 \\
  32 & 19.6983383(700)   & 227.990027(810)  & 40.2 (8) & 4.84164 (301) &  359.3 &   16.9 &  435.095 \\
  33 & 15.8791602(704)   & 183.786576(815)  & 39.9 (8) & 1.55301 (303) &  348.0 &   16.0 &  433.329 \\
  34 & 17.3722601(750)   & 201.067825(868)  & 37.5 (8) & 5.34496 (322) &  344.4 &   15.0 &  432.464 \\
  35 & 20.5323997(728)   & 237.643515(843)  & 38.6 (8) & 4.42699 (313) &  328.3 &   16.7 &  429.857 \\
  36 & 18.6159636(756)   & 215.462542(875)  & 37.2 (8) & 2.16777 (325) &  326.1 &   14.6 &  429.025 \\
  37 & 17.5479304(790)   & 203.101046(914)  & 35.6 (8) & 6.03357 (340) &  322.8 &   14.2 &  426.606 \\
  38 & 19.5626027(638)   & 226.419013(738)  & 44.0 (8) & 0.31765 (274) &  265.0 &   18.6 &  419.077 \\
  39 & 20.9774212(821)   & 242.794227(950)  & 34.2 (8) & 3.05581 (353) &  265.8 &   14.6 &  418.292 \\
  40 & 15.0815641(781)   & 174.555140(904)  & 35.9 (8) & 6.00459 (336) &  265.0 &   14.8 &  417.562 \\
  41 & 15.5735859(820)   & 180.249837(949)  & 34.3 (8) & 5.96530 (353) &  261.5 &   13.8 &  415.125 \\
  42 & 15.2514736(410)   & 176.521685(475)  & 68.6 (8) & 2.26019 (176) &  255.6 &   27.9 &  414.361 \\
  43 & 17.8087368(819)   & 206.119639(948)  & 34.3 (8) & 4.10416 (352) &  250.6 &   13.4 &  413.664 \\
  44 & 23.7935704(922)   & 275.388546(1067) & 30.5 (8) & 5.55366 (396) &  242.0 &   14.2 &  411.587 \\
  45 & 19.3525019(957)   & 223.987291(1108) & 29.3 (8) & 1.28361 (412) &  218.3 &   12.0 &  408.576 \\
  46 & 19.6436014(903)   & 227.356498(1045) & 31.1 (8) & 0.07760 (388) &  219.1 &   13.1 &  408.021 \\
  47 & 24.4158193(944)   & 282.590501(1093) & 29.7 (8) & 2.84486 (406) &  209.7 &   13.9 &  406.871 \\
  48 & 12.7201671(1007) &  147.224156(1166) & 27.9 (8) & 6.15249 (433) &  211.1 &   12.3 &  397.691\\
  49 & 23.4292388(1024) &  271.171745(1185) & 27.4 (8) & 5.73862 (440) &  206.0 &   12.8 &  396.946\\
  50 & 15.1877607(1072) &  175.784267(1241) & 26.2 (8) & 2.98303 (461) &  203.4 &   10.7 &  395.936\\
\enddata 
\end{deluxetable}
%-table-%


\begin{thebibliography}{}

\bibitem[Baglin(2006)]{corot} Baglin, A. 2006, in The CoRoT Mission: Pre-Launch Status, ed. M. Fridlund, A. Baglin, J. Lochard, \& L. Conroy, ESA Publications Division, Noordwijk, ESA SP, 1306
%\bibitem[Balona(2014)]{Balona2014} Balona, L. A. 2014, \mnras, 439, 3453–3460
%\bibitem[Barnes(2009)]{Barnes2009} Barnes, J. W. 2009, \apj, 705, 683
\bibitem[Breger(1979)]{Breger1979} Breger, M. 1979, \pasp, 91, 5-26
%\bibitem[Breger et al.(1993)]{Breger1993} Breger, M., Stich, J., Garrido, R., et al. 1993, \aap, 271, 482
\bibitem[Breger et al.(1999)]{Breger1999} Breger, M., Pamyatnykh, A.~A., Pikall, H., et al.1999, \aap, 341, 151
\bibitem[de Bruijne(2012)]{deBruijne2012} de Bruijne, J. H. J. 2012, \apjs, 341, 31-41
\bibitem[Chapellier \& Mathias(2013)]{Chapellier2013} Chapellier, E., \& Mathias, P. 2013, \aap, 556, A87
\bibitem[Creevey et al.(2009)]{Creevey2009} Creevey, O. L., Uytterhoeven, K., Mart{\'{\i}}n-Ruiz, S., et al. 2009, \aap, 507, 901-910
\bibitem[Garc\'{\i}a Hern\'andez et al.(2009)]{GH09} Garc{\'{\i}}a Hern{\'a}ndez, A., Moya, A., Michel, E., et al. 2009 \aap, 506, 79G
\bibitem[Garc{\'{\i}}a Hern{\'a}ndez et~al.(2013a)]{GH13a} Garc{\'{\i}}a Hern{\'a}ndez, A., Moya, A., Michel, E., et al. 2013, \aap, 559, A63
\bibitem[Garc{\'{\i}}a Hern{\'a}ndez et~al.(2013b)]{GH13b} Garc{\'{\i}}a Hern{\'a}ndez, A., Pascual-Granado, J., Grigahc{\`e}ne, A., et al. 2013, in \apss\ Proceedings, Vol.~31, Stellar Pulsations: Impact of New Instrumentation and New Insights, ed. {J.~C.~{Su{\'a}rez}, R.~{Garrido}, L.~A.~{Balona} \& J.~{Christensen-Dalsgaard}}, 61-65
\bibitem[Hambleton et al.(2013)]{Hambleton2013} Hambleton, K. M., Kurtz, D. W., Pr{\v s}a, A., et al. 2013, \mnras, 434, 925-940
\bibitem[Handler et al.(1997)]{Handler1997} Handler, G., Pikall, H., O'Donoghue, D., et al. 1997, \mnras, 286, 303
\bibitem[Hareter et al.(2014)]{Hareter2014} Hareter, M., Papar\'o, M., Weiss, W., et al. 2014, \aap, 567, A124
\bibitem[Hinkley et al.(2011)]{Hinkley2011} Hinkley, S., Monnier, J. D., Oppenheimer, et al. 2011, \apj, 726, 104
\bibitem[Ibano{\v g}lu et al.(2006)]{Ibanoglu2006} Ibano{\v g}lu, C., Soydugan, F., Soydugan, E., et al. 2006, \mnras, 373, 435-448
\bibitem[Kjeldsen(2008)]{Kjeldsen2008} Kjeldsen, H., Bedding, T. R., \& Christensen-Dalsgaard, J. 2008, \apj, 683, L175-L178
\bibitem[Koch et al.(2010)]{kepler} Koch, D. G., Borucki, W. J., Basri, G., et al. 2010, ApJ, 713, L79
%\bibitem[Kuschnig et al.(1997)]{Kuschnig1997} Kuschnig, R., Weiss, W. W., Gruber, R., et al. 1997, \aap, 328, 544-550
\bibitem[Lehmann et al.(2013)]{Lehmann2013} Lehmann, H., Southworth, J., Tkachenko, A., et al. 2013, \aap, 557, A79
\bibitem[Ligni\`eres \& Georgeot(2008)]{Lignieres2008} Ligni\`eres, \& Georgeot, B. 2008, Phys. Rev. E 78, 016215
\bibitem[Ligni\`eres \& Georgeot(2009)]{Lignieres2009} Ligni\`eres, \& Georgeot, B. 2009, \aap, 500, 1173
\bibitem[Ligni\`eres et al.(2006)]{Lignieres2006} Ligni\`eres, F., Rieutord, M., \& Reese, D. 2006, \aap, 455, 607
\bibitem[MacGregor et al.(2007)]{MacGregor2007} MacGregor, K. B., Jackson, S., Skumanich, A., et al. 2007, \apj, 663, 560-572
\bibitem[Maeder(2009)]{Maeder2009} Maeder, A. 2009, Formation and Evolution of Rotating Stars (\aap\ Library; Springer Berlin Heidelberg)
\bibitem[Maceroni et al.(2014)]{Maceroni2014} Maceroni, C., Lehmann, H., da Silva, R., et al. 2014, \aap, 563, A59
\bibitem[Martin(2003)]{Martin2003} Martin, S. 2003, Survey in search of variable stars in open clusters, Interplay of Periodic, Cyclic and Stochastic Variability in Selected Areas of the H-R Diagram. Ed. by C. Sterken, ASP Conf. Ser, Astronomical Society of the Pacific, San Francisco, 292, 59
\bibitem[Mart{\'{\i}}n-Ruiz et al.(2005)]{MartinRuiz2005} Mart{\'{\i}}n-Ruiz, S., Amado, P. J., Su\'arez, J. C. 2005, \aap, 440, 711-714
\bibitem[Monnier et al.(2010)]{Monnier2010} Monnier, J. D., Townsend, R. H. D., Che, X. et al. 2010, \apj, 725, 1192
\bibitem[Montgomery \& Odonoghue(1999)]{Montgomery1999} Montgomery, M. H., \& Odonoghue, D. 1999, Delta Scuti Star Newsletter, 13, 28
\bibitem[Ouazzani et al.(2015)]{Ouazzani2015} Ouazzani, R. M., Roxburgh, I. W., \& Dupret, M. A. 2015, \aap, in press
\bibitem[Pascual-Granado et al.(2015)]{PascualGranado2015} Pascual-Granado, J., Garrido, R., \& Su\'arez, J. C. 2015, \aap, 575, 78
\bibitem[Pasek et al.(2012)]{Pasek2012} Pasek, M., Ligni\`eres, F., Georgeot, B., et al. 2012, \aap, 546, A11
\bibitem[Rauer et al.(2014)]{Rauer2014} Rauer, H., Catala, C., Aerts, C., et al. 2014, Exp. Astron., 38, 249-330
\bibitem[Reegen(2007)]{Reegen2007} Reegen, P. 2007, \aap, 467, 1353–1371
\bibitem[Reese et al.(2008)]{Reese2008} Reese, D. R., Ligni\`eres \& F., Rieutord, M. 2008, \aap, 481, 449-452
\bibitem[Reese et al.(2009)]{Reese2009} Reese, D. R., MacGregor, K. B., Jackson, S., et al. 2009, \aap, 506, 189-201
\bibitem[Royer et al.(2007)]{Royer2007} Royer, F., Zorec, J. \& G\'omez, A. E., 2007, \aap, 463, 671
%\bibitem[Seager \& Mall\'en-Ornelas(2003)]{Seager2003}  Seager, S. \& Mall\'en-Ornelas, G., 2003, \apj, 585, 1038-1055
\bibitem[da Silva et al.(2014)]{daSilva2014} da Silva, R., Maceroni, C., Gandolfi, D., et al. 2014, \aap, 565, A55
\bibitem[Su\'arez et al.(2014)]{Suarez2014} Su\'arez, J. C., Garc{\'{\i}}a Hern\'andez, A., Moya, A., et al. 2014, \aap, 563, A7
\bibitem[Tassoul(1980)]{Tassoul1980} Tassoul, M. 1980, \apjs, 43, 469-490
\bibitem[Walker et al.(2003)]{most} Walker, G., Matthews, J., Kuschnig, R., et al. 2003, \pasp, 115, 1023–1035


\end{thebibliography}
\end{document}